\documentclass[useAMS,usenatbib,usegraphicx]{mn2e}

\newcommand{\snd}{secondary}

\newcommand{\pri}{primary}
\newcommand{\rv}{radial velocity}
\newcommand{\rvs}{radial velocities}

\newcommand{\ks}{km s$^{-1}$}
\newcommand{\mbs}{maximum blueshift}

\newcommand{\wtd}{white dwarf}
\newcommand{\ha}{H$\alpha$}
\newcommand{\hb}{H$\beta$}
\newcommand{\hg}{H$\gamma$}
\newcommand{\heo}{HeI $\lambda$4471}
\newcommand{\het}{HeII $\lambda$4686}
\newcommand{\el}{emission line}
\newcommand{\els}{emission lines}

\newcommand{\bbc}{broad-base component}

\usepackage{times}
\usepackage{mathptm}
\begin{document}


\title[Outbursts of EX Hydrae Revisited]{Outbursts of EX Hydrae Revisited}
\author[N. Mhlahlo, D.A.H. Buckley, V. Dhillon, S. Potter, B. Warner and P. Woudt]{N. Mhlahlo$^{1}$\thanks{E-mail:
nceba@maia.saao.ac.za}, D.A.H. Buckley$^{2}$, V.S. Dhillon$^{3}$, S.B. Potter$^{2}$, B. Warner$^{1}$  and \newauthor P.A. Woudt$^{1}$ \\
$^{1}$Department of Astronomy, University of Cape Town, Rondebosch 7700, Cape Town, South Africa \\
$^{2}$South African Astronomical Observatory, Observatory 7935, Cape Town, South Africa \\
$^{3}$Physics and Astronomy Department, University of Sheffield, Sheffield, S3 7RH, UK}


\maketitle

\label{firstpage}

\begin{abstract}
We present optical spectroscopy of EX Hya during its 1991 outburst.  This outburst is characterised by strong irradiation of the front face of the \snd\ star by the \wtd, an overflowing stream which is seen strongly in \het\ and by a dip in the light curves, which extends from 0.1-0.6 in the binary and spin phases.  Strong irradiation of the accretion curtain and that of the inner regions of the disc led to strong emission of \het\ and to the suppression of the \hg\ and \hb\ emission.

Disc overflow was observed in quiescence in earlier studies, where the overflow stream material was modulated at high velocities close to 1000 \ks.  In outburst, the overflowing material is modulated at even higher velocities ($\sim1500$ \ks).  These are streaming velocities down the field lines close to the \wtd.
Evidence for material collecting near the outer edge of the disc and corotating with the accretion curtain was observed.  In decline, this material and the accretion curtain obscured almost all the emission near binary phase 0.4, causing a dip.  The dip minimum nearly corresponds with spin pulse minimum.  This has provided additional evidence for an extended accretion curtain, and for the corotation of material with the accretion curtain at the outer edge of the disc.  From these observations we suggest that a mechanism similar to that of Spruit \& Taam, where outbursts result due to the storage and release of matter outside the magnetosphere, triggers the outbursts of EX Hya.  This is followed by the irradiation of the \snd\ star due to accretion induced radiation.  
\end{abstract}


\section{Introduction}

EX Hya is an eclipsing binary system and an Intermediate Polar (IP), a sub-class of magnetic Cataclysmic Variable Stars (mCVs) where a white dwarf star accretes material from a late-type main sequence star, or the \snd, as the two stars orbit about their common centre of mass under the influence of their mutual gravitation.  

EX Hya has a spin period ($\sim$67.03 minutes) which is about $2/3$ its orbital period (98.26 minutes \cite{mum67,hel87}) and has its accretion curtains extending to the outer edge of the accretion disc (hereafter the accretion ring or the ring) near the Roche lobe radius \citep{kin99,wyn00,nor04,bel02,bel05,mhl06}.  Recently, \cite{mhl06} showed that near the Roche lobe radius the accretion curtains corotate with part of the accretion ring and a combination of stream and ring accretion (ring overflow) was explained well by the observations. 
 
EX Hya is one of the few IPs that undergo brief outbursts, nearly once every 1.5 years \citep{hel00a}.  Previous spectral studies of EX Hya in outburst have been characterised mainly by two events: the development of a \bbc\ of the \els\ on the night of outburst, possibly modulated at the orbital period \citep{hel89b}, and the significant decrease, or even absence, of the modulation at the spin pulse \citep{bon87,hel89b}.  

Studies done by \cite{hel89b} during the 1987 outburst of EX Hya revealed a broad base component (the `base excursion') that appeared to be modulated at the orbital period.  The `base excursion' was phased with \mbs\ at orbital phase $\phi_{98} \sim0.4$ and was interpreted as being caused by the stream of material overflowing the initial impact with the disc and free-falling onto the magnetosphere of the \wtd.  These results were confirmed during an outburst of EX Hya in 2000 where an X-ray beat pulse was discovered and eclipse of the overflow stream material by the \snd\ was observed \citep{hel00a}.  

During the 1991 outburst a dip feature was observed in the light curves in decline, where almost all the emission was obscured around phase $\phi_{67} \sim0.4$ \citep{buc92}.  No interpretation was provided for this dip.

Five other IPs that have been reported to undergo outbursts are XY Ari \citep{hel97}, GK Per \citep{sab83,wat85,mor96}, TV Col \citep{szk84,hel93}, V1223 Sqr \citep{vap89} and YY Dra \citep{szk02}.
XY Ari, GK Per and YY Dra are said to have their outbursts caused by an instability in the disc \citep{hel97} while the other three IPs, TV Col, V1223 Sqr and EX Hya seem to pose problems for the disc instability model.  It has been suggested that the outbursts in these three stars are a result of increased mass transfer from the \snd\ \citep{hel93,hel89b,hel00a}.  An alternative is that the enhanced mass transfer could be a result of increased irradiation of the \snd, which in turn is a consequence of a disc instability \citep{hel00a}. 
 
With these questions in mind, we present the outburst data of EX Hya and the analysis in Sections~\ref{sec:obsred} - \ref{o:rvcurve}, we investigate the cause of the dip seen in the light curves in Section~\ref{sec:dips} and we discuss and interprete the results in Section~\ref{sec:odisc}.

\section{\bf Observations And Data Reduction}

\label{sec:obsred} 
In this section we present spectroscopic data obtained in 1991 at the South African Astronomical Observatory (SAAO) when EX Hya was in outburst.

\begin{table}
\begin{center}
\small
\caption{\small Table of spectroscopic observations during the 1991 outburst.  The column \textbf{No. of Hrs} denotes the number of hours of observations, \textbf{Spec.} the number of spectra obtained and \textbf{State} is either outburst (o), decline (d) or quiescence (q).  A grating with a resolution of 1200 mm$^{-1}$ was used and covered a wavelength range of 4000-5080~\AA.} \label{tabo:observ}
\begin{tabular}{|cccccc|}  \hline
\multicolumn{1}{|c|}{\textbf{Date}} &
\multicolumn{1}{c|}{\textbf{HJD (start)}} &
\multicolumn{1}{c|}{\textbf{No. of Hrs}} &
\multicolumn{1}{c|}{\textbf{Spec.}} &
\multicolumn{1}{c|}{\textbf{State}}                   \\ \hline
24-04-91 & 2448371.3884097 &  $3.28$  &  $90$  & q \\  
25-04-91 & 2448372.3603850 &  $2.93$  &  $72$  & q  \\   
27-04-91 & 2448374.2808828 &  $5.45$  &  $160$ & o \\ 
28-04-91 & 2448375.2925521 &  $5.49$  &  $150$ & d \\ 
29-04-91 & 2448376.2402973 &  $7.00$  &  $100$ & q \\ \hline
\end{tabular}   
\end{center}
\end{table} 
This occured on 27/28 April 1991 and was observed using the SAAO 1.9-m telescope with the Reticon photon counting system (RPCS) detector on the Cassegrain spectrograph by \cite{buc91}.  
A wavelength range  of 4000 - 5080 \AA~was covered at a spectral resolution of $\Delta \lambda \sim$1.2 \AA~and at a time resolution of 100 - 120 s.  
The spectrograph slit width was 250 $\mu$m ($\sim$1.5 arcsecs).  Wavelength calibration exposures were taken using a CuAr arc lamp.  
Five nights of observations (24, 25, 27, 28 and 29 April 1991) were covered and in total 572 spectra were obtained.  The observing log is given in Table~\ref{tabo:observ} showing the starting times of the observations, the data length and the number of spectra obtained on each night (the data of 24, 25 and 29/04/91 were presented in \cite{mhl06}).  
Following the wavelength-calibration and sky-subtraction, the data were flux calibrated using the spectra of the standard star LTT3864.

Preliminary analysis of the data was done and initial results were published by \cite{buc91} and \cite{buc92}. The data were archived and have been retrieved for a more detailed analysis, presented here.

\section{\bf The visual light curves}

The 1991 outburst of EX Hya started on 27/28 April, with the system rising from $V \sim$13 mag to $V \sim$10 mag.  The outburst lasted for $\sim$1-2 days, during which EX Hya increased in brightness by $\sim$3 magnitudes.  
\begin{figure}
\begin{center}
\includegraphics[width=60mm]{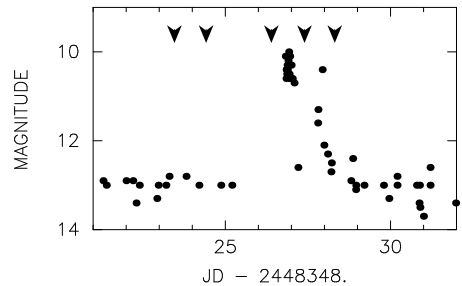}
\caption{\small Visual light curves of EX Hya before, during and after the 1991 outburst.  The data were supplied by the VSS RASNZ.  The arrows show the times of our observations.}
\label{l:visual}
\end{center}
\end{figure}
Figure~\ref{l:visual} shows the visual light curves, with the times of our spectroscopic observations represented by arrows; they represent an average of start and end times of our observations of each night.  The visual light curve data were supplied by the Variable Star Section of the Royal Astronomical Society of New Zealand (VSS RASNZ).
\section{The Emission Line Profiles}
\begin{figure}
\begin{center}
\includegraphics[width=70mm]{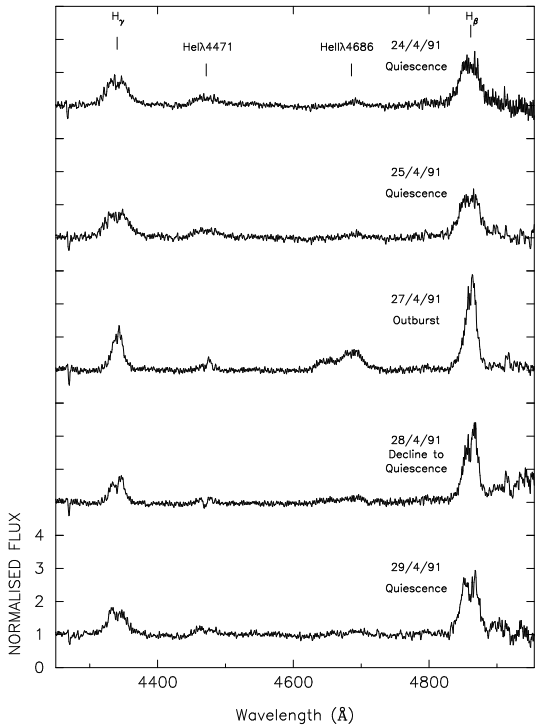}
\caption{\small The summed spectra of EX Hya in quiescence, in outburst and during decline from outburst.  The spectra have been normalised by the continuum and all panels have the same scale.}
\label{a:avx7}
\end{center}
\end{figure}

Figure~\ref{a:avx7} shows the summed spectra of EX Hya during rise to outburst, in outburst and during decline to quiescence.  The spectra have been normalised by the continuum to aid comparison.
Prominent peaks in the spectra are those of the \hb\ and \hg.  
As was reported by \cite{buc91}, the strengthening of all the emission lines is evident during outburst.  The HeII $\lambda$4686 line, which was weak in quiescence, emerges stronger and broad in outburst, while the HeI $\lambda$4471 shows a strong central absorption component on the night following outburst. 
The \hb\ and \hg\ \els\ are double-peaked and seem to be broader in quiescence than in outburst.  

\begin{figure}
\begin{center}
\includegraphics[width=60mm]{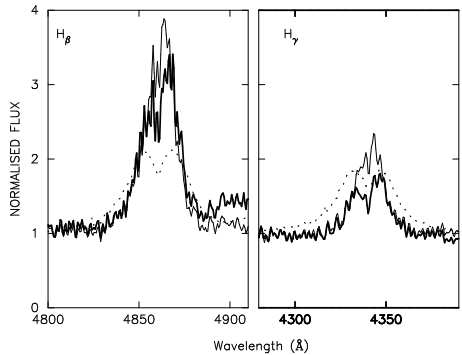}
\caption{\small The \hb\ and \hg\ summed spectra in outburst (light line), in decline (dark line) and in quiescence (dashed line) are plotted superimposed.  All the data are normalised by the continuum.}
\label{s:superimp}
\end{center}
\end{figure}

The line profile widths showed a significant decrease during outburst, by $\sim$33\% for \hb, $\sim$50\% for \hg\ and $\sim$50\% for \heo\ (not shown), when compared to the line widths in quiescence of 1991; and by $\sim$50\% for \hb\ and \hg\ and by $\sim$55\% for \heo\ (not shown), when compared with the line widths in quiescence of 2001 (see Figure~\ref{s:superimp}).  An increase in the widths of the emission lines, with respect to outburst, by $\sim$20\% in \hb\ and \hg\ (Figure~\ref{s:superimp}) and by $\sim$33\% in \heo\ (not shown), was observed during decline.
We find a full width at half maximum (FWHM) of $\sim 38$ \AA~for both \hb\ and \hg\ in quiescence and $\sim$20 \AA~for both emission lines in outburst.  
The Full Width at Zero Intensity (FWZI) was found to be $\sim$ 3000 km s$^{-1}$ for \hb\ and $\sim$ 3400 km s$^{-1}$ for \hg\ in outburst, compared to that of $\sim$ 7000 km s$^{-1}$ ($\sim$ 180 \AA) in quiescence, for both the \hb\ and \hg\ \els\ (see also \cite{hel87,hel89b} for similar results).
The emission line components are identified in the trailed spectra in Section~\ref{sec:o-orbtomo}.
\section{The Orbital Radial Velocities}
\label{rv:oorv}

The \rvs\ in outburst of 1991 were obtained using the Gaussian Convolution Scheme (e.g. Shafter \& Szkody 1984; Shafter 1985).  The GCS method convolves each spectrum with two identical Gaussian, one in the red wing and one in the blue wing.  The separation between the two Gaussians is 2$\alpha$ (see also \cite{mhl06}).  The \rvs\ were measured from the wings and the core of the \hb, \hg\ and HeII $\lambda4686$ emission lines from every individual spectrum, and the data were used to determine the periods from Discrete Fourier Transforms (DFTs); \citep{dee75,kur85}.  

\subsection{Period Searches in the Line Wings}

\begin{figure}
\begin{center}
\includegraphics[width=80mm]{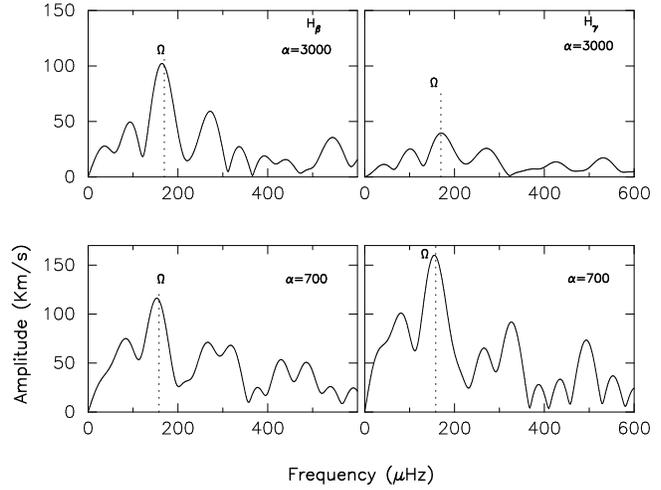}
\caption{\small The amplitude spectra of the \hb\ and \hg\ radial velocities during outburst.  The broad-base component and the narrow-peak are modulated at the orbital period.}
\label{rv:27hbg-as}
\end{center}
\end{figure}

Figure~\ref{rv:27hbg-as} shows the amplitude spectra of the \hb\ and \hg\ radial velocities (\het\ shows similar results).
\begin{figure}
\begin{center}
\includegraphics[width=75mm]{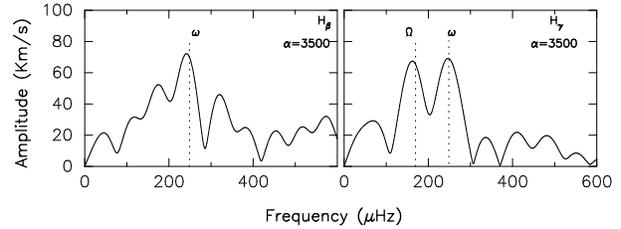}
\caption{\small The amplitude spectra of the \hb\ and \hg\ radial velocities during decline.  The spin pulse emerges during decline.}
\label{rv:2891hbg-s}
\end{center}
\end{figure}
Velocities in the line wings are dominated by the 98-min orbital velocity variation.  The $\Omega$ modulation, where $\Omega$ denotes the orbital frequency of the system, is also present in the line cores of \hb\ and \hg.

The spin pulse, which is not detectable during outburst, is present in the line wings during decline (Figure~\ref{rv:2891hbg-s}).  The \hb\ and \hg\ DFTs do not show any variation in the core in decline and the \heo\ DFT shows a variation at the second harmonic of the orbital frequency (not shown). 

\section{\bf Orbital Variations of the Emission Lines in Outburst}
\label{sec:ovemo}

All the \rv\ data were phase-folded on the orbital ephemeris of \cite{hel92} which is defined by the zero phase being mid-eclipse.  For the \rvs, \mbs\ occurs at phase 0.75 when viewing is perpendicular to the line of centres.  50 phase intervals were used.

\subsection{Doppler Tomograms and Trailed Spectra}
\label{sec:o-orbtomo}

\begin{figure*}
\begin{center}
\includegraphics[width=100mm]{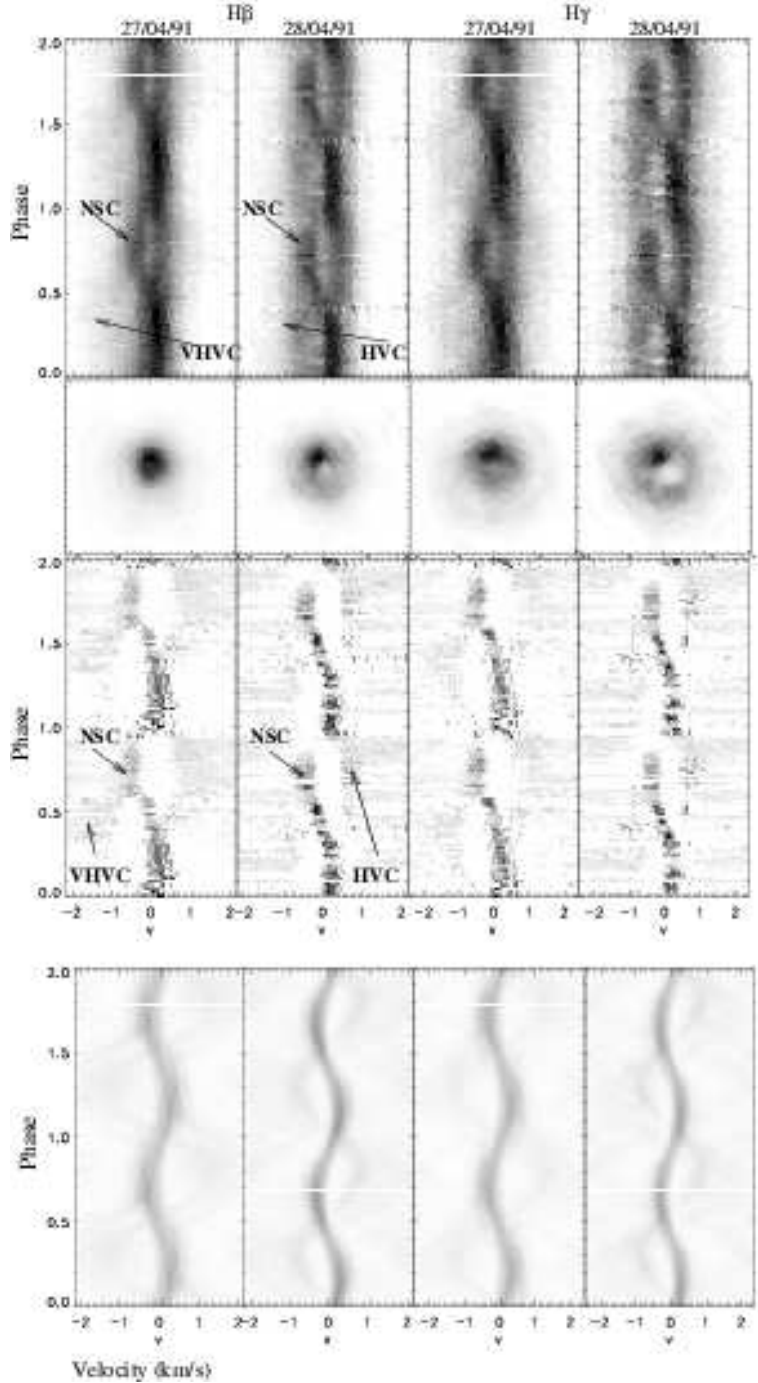}
\caption{\small \hb\ and \hg\ trailed spectra in outburst and during decline are shown (top panels).  MEM orbital tomograms are shown in the second panels with the average-subtracted trailed spectra shown in the third panels.  The emission line components are identified for \hb, before and after subtracting the average (the \hg\ line shows similar components). All the data are plotted on the same velocity scale.  The reconstruction is shown in the bottom.  The velocity units along the x-axis are 10$^{3}$ \ks.}
\label{o:hbgidltrl}
\end{center}
\end{figure*}
\begin{figure*}
\begin{center}
\includegraphics[width=95mm]{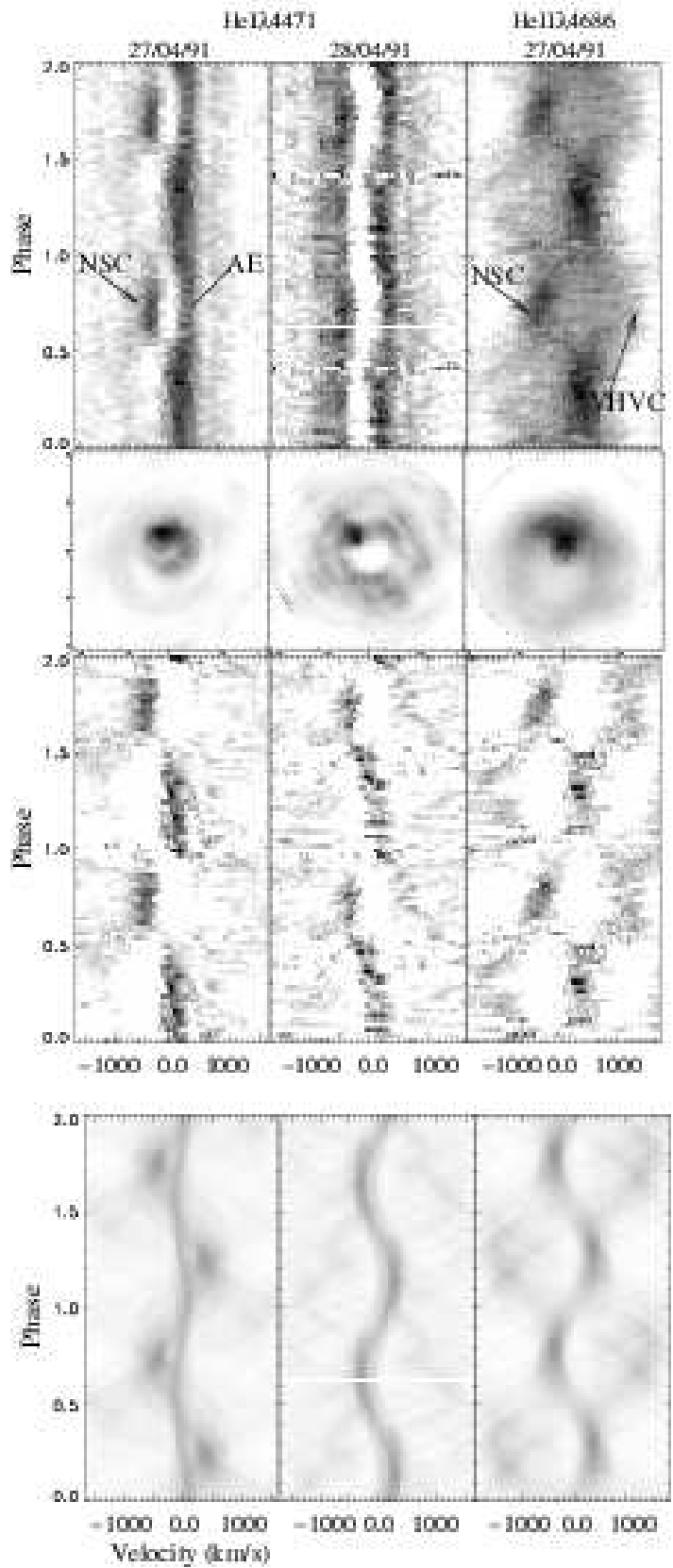}
\caption{\small \heo\ trailed spectra in outburst and during decline, and HeII $\lambda4686$ trailed spectra in outburst are displayed.  Orbital tomograms are shown in the second panels.  The third panels show the trailed spectra after average-subtraction and the bottom panels show the reconstruction.  The data were folded on the orbital period.  The NSC and the `arc emission' (AE) in \heo\ are shown.  In \het, the dashed line shows the orbital motion of the NSC, and the VHVC is identified in the line wings.}
\label{o:he12idltrl}
\end{center}
\end{figure*}
\begin{figure*}
\begin{center}
\includegraphics[width=150mm]{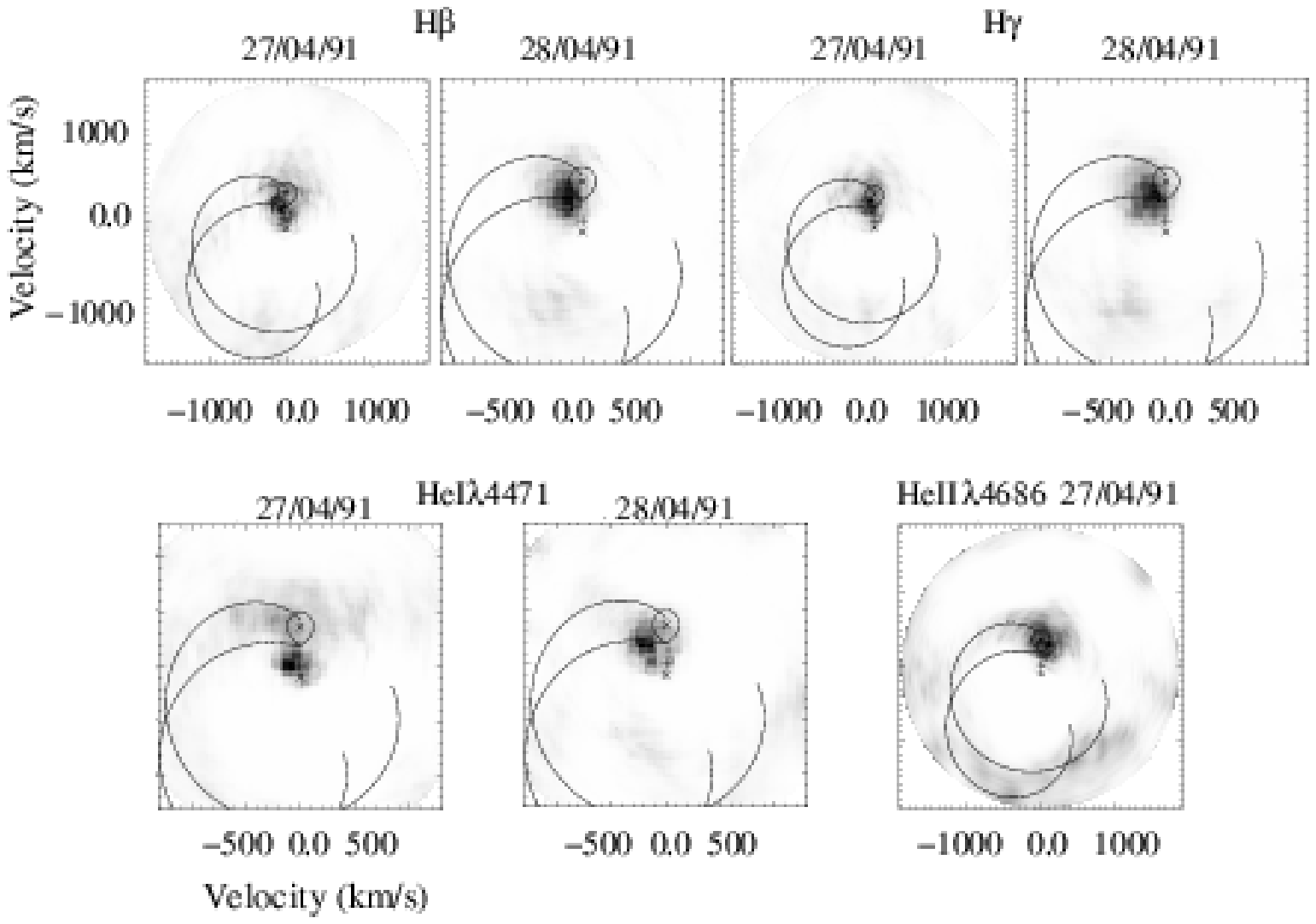}
\caption{\small \hb\ and \hg\ average subtracted orbital tomograms in outburst (first and third top panels) and during decline (second and fourth top panels) are displayed.  The tomograms in decline are expanded since the emission observed at high velocities in outburst (near V$_{y}\sim1500$ \ks) has moved to lower velocities (near V$_{y}\sim800$ \ks) in decline.  \heo\ and \het\ tomograms are shown in the bottom panels.}
\label{o:avsubmaps}
\end{center}
\end{figure*}

The Maximum Entropy Method (MEM) \citep{spr98} was used to construct the orbital Doppler tomograms, using the phase binned data.  For each spectrum an average was taken and subtracted from the spectrum to emphasize the bright spot and secondary star emission.  The \hb, \hg, \heo\ and \het\ tomograms and trailed spectra are shown in Figures~\ref{o:hbgidltrl} and \ref{o:he12idltrl}, and the average-subtracted tomograms are shown in Figure~\ref{o:avsubmaps}, in outburst and in decline.  

In the trailed spectra, a narrow S-wave component (NSC) can be detected in all the \els, before and after subtracting the average.  
The NSC shows \mbs\ near phase $\phi_{98} \sim 0.75$ as expected from the motion of the \snd, and an amplitude velocity of $\sim400$ \ks\ (obtained from fitting a gaussian to the \hb\ NSC between phases 0.5-1.0).  The latter is slightly greater than but close to the K-velocity amplitude of the \snd\ of $K_{2}=360\pm35$ kms$^{-1}$ \citep{bel03,put03,beu03}.  These observations suggest that the NSC originated from the \snd.  
The NSC did not originate from the bright spot since it shows a blueshift during binary phases $\phi_{98} \sim 0.61-0.85$ whereas the NSC resulting from the bright spot is expected to show zero radial velocity emission around these phases \citep{bel05,mhl06}.  The NSC emission moves (from the Roche lobe in outburst) to near the bright spot velocities (slightly between the bright spot and the Roche lobe) in decline.

A \bbc\ (see \cite{hel89b}) or a very high velocity component (hereafter the VHVC since its velocity amplitude is higher than that of the high velocity component (HVC) seen in quiescence - see \cite{mhl06} and \cite{ros87} - and in decline) can be observed in the trailed spectra in outburst.  The VHVC is modulated with a velocity of $\pm1500$ \ks\ and is much enhanced in \het.  
In \cite{hel89b}, it was not certain if this component (which they observed in \ha) varied with the 98-min cycle of the binary since their data covered only 1.2 orbital cycles.  Our 1991 outburst data covered 3.3 orbital cycles and are, at least, sufficient to conclude that the motion of this component is consistent with the 98-min orbital period.
 
The NSC and the VHVC are superimposed at phases $\phi_{98}\sim0.1-0.4$ and $\phi_{98} \sim0.6-0.9$ for \hb, \hg\ and \heo, and at phases $\phi_{98}$ $\sim 0.25-0.5$ and $\phi_{98}$ $\sim 0.6-0.85$ for the \het\ \el\ (Figures~\ref{o:hbgidltrl} and \ref{o:he12idltrl}).  A similar effect was observed in quiescence \citep{mhl06} near phase $\phi_{98}\sim0.25$.  

Outburst tomograms show strong emission at and around the Roche lobe (seen more clearly in Figure~\ref{o:avsubmaps}) for all the emission lines (except for \heo\ where strong emission is seen near V$_{x,y}=0$ than at the Roche lobe), most especially in \het, further suggesting that the NSC was mainly caused by the \snd\ star.  

The tomograms (more especially \het) suggest that the emission which lies on the `far' side (nearly on the opposite side of the disc, relative to the front face of the \snd), behind the \wtd, observed at high velocities ($\sim 1500$ \ks), was responsible for the VHVC (Figure~\ref{o:avsubmaps}).  This emission moves to lower velocities ($\sim 800 - 900 $ \ks) in decline and can be associated with the HVC seen in the trailed spectra around 900 \ks.  This HVC is reminiscent of that observed in the tomograms in quiescence \citep{mhl06} on the bottom left quadrant and near the opposite side of the disc, at velocities near 800 -- 900 \ks, which originated from the overflow stream.

The \hb\ and \hg\ NSC is less bright at phases $\phi_{98}\sim0.6-0.9$ and this could be the effect of Vertically Extended Material (VEM) which was observed in quiescence obscuring emission around these phases.  There was no simultaneous photometry obtained, though, and so observed changes in brightness of the NSC and other components are not secure.

It is important to note that during outburst the relative amount of the accretion ring line emission increased and that most of the emission came from the outer edge of the ring.  
This is, however, not the case in the \het\ \el\ which, interestingly, shows stronger disc emission even at high velocities than the other emission lines (Figure~\ref{o:he12idltrl}).  This emission is clearly observed on the `far' side (relative to the Roche lobe), behind the \wtd, after subtracting the average which is supposed to remove most of the symmetric emission originating mainly from the disc.  

Interestingly, the \heo\ tomogram in outburst (Figure~\ref{o:he12idltrl}) shows an enhancement of emission curving away from the \snd\ along the outer edge of the disc, from $\phi_{98}= 0$ (12 o'clock position) to 0.5 (near the 6 o'clock position), in the clockwise direction (hereafter the `arc emission' or AE).  Between phases $\phi_{98}= 0.0-0.6$, the \heo\ trailed spectrum (before subtracting the average) shows strong redshifted emission, resulting from the superposition of the NSC and the `arc emission'.  At these phases almost all the blueshifted emission is absorbed.  Between phases $\phi_{98}= 0.6-0.9$, the NSC is blueshifted while the `arc emission' is still redshifted.  This suggests that the material responsible for the `arc emission' is drifting away from the observer in the radial direction at all orbital phases.
During decline, the central absorption feature in \heo\ becomes stronger, and the tomogram nolonger shows enhanced emission at the outer edge of the disc.

\subsection{The Radial Velocity Curve}
\label{o:rvcurve}
The \rv\ data were fitted on the orbital frequency using a fitting function of the form, 
\begin{figure}
\begin{center}
\includegraphics[width=80mm]{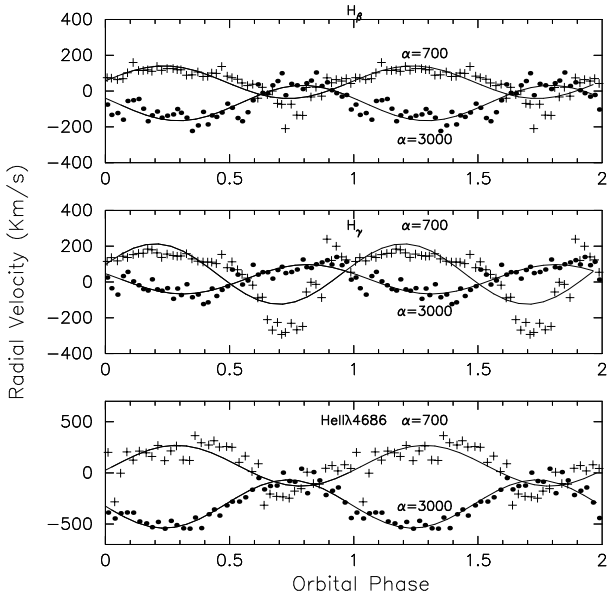}
\caption{\small The \hb, \hg\ and HeII $\lambda4686$ radial velocities plotted as a function of the orbital phase.  The VHVC (dots) is nearly antiphased to the narrow-peak component (crosses).  The solid line represents a fit to the data.}
\label{rv:2791hbghe2-3t7h1t2}
\end{center}
\end{figure}
\begin{equation}
V(t) = \gamma + K {\rm cos} (2\pi*\Omega*t + \phi) 
\label{eq:rvfit1}
\end{equation}
where $t$ are the times in Heliocentric Julian Days (HJDs).  
Figure~\ref{rv:2791hbghe2-3t7h1t2} shows a comparison between the \hb, \hg\ and \het\ NSC and the VHVC.  The top panel shows the \hb\ VHVC having maximum redshift at $\phi_{98} \sim 0.79$ and maximum blueshift at $\phi_{98} \sim 0.29$ while the NSC is phased with maximum blueshift at $\phi_{98}$ $\sim 0.74$ and lags the VHVC by $\phi_{98} \sim 0.45$ ($\sim 162^{\circ}$).  
The \hg\ \rv\ curve (middle panel) shows the VHVC having maximum redshift at $\phi_{98} \sim 0.81$ and maximum blueshift at $\phi_{98} \sim 0.31$.  The NSC is phased with maximum blueshift at $\phi_{98}$ $\sim 0.7$ and lags the VHVC by $\phi_{98} \sim 0.39$ ($\sim 140^{\circ}$).  

The $\alpha=700$ km s$^{-1}$ curve has an interesting shape as it shows very negative velocities ($\sim -350$ \ks) around $\phi_{98} \sim 0.7$, indicating that there is streaming material from the accretion stream joining the disc at the bright spot.  It is curious that \hb\ does not show a similar effect.
The HeII $\lambda4686$ \rv\ curve (bottom panel) shows the VHVC having maximum redshift at $\phi_{98} \sim 0.74$ and maximum blueshift at $\phi_{98} \sim 0.24$, and the NSC phased with maximum blueshift at $\phi_{98}$ $\sim 0.79$ and lagging the VHVC by $\phi_{98} \sim 0.55$ ($\sim 198^{\circ}$), all in agreement with the behaviour of these components in the trailed spectra. 
\section{The 'Dip'}
\label{sec:dips}
\begin{figure}
\begin{center}
\includegraphics[width=80mm]{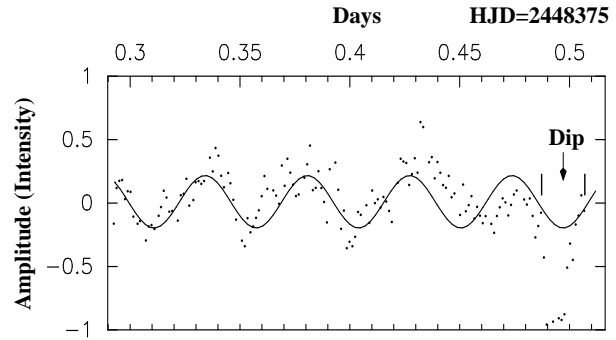}
\caption{A normalised light curve of the \heo\ emission line showing a variation at the spin cycle, and the dip.  The solid line represents a fit to the data.}
\label{o:diplcfit}
\end{center}
\end{figure}
\begin{figure}
\begin{center}
\includegraphics[width=80mm]{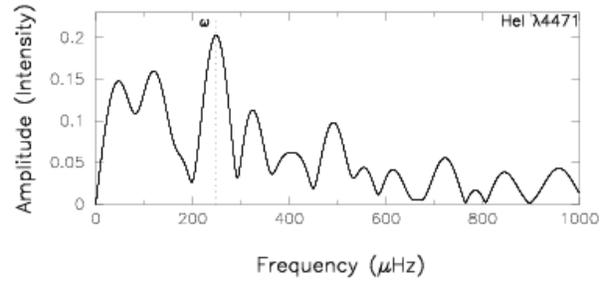}
\caption{\heo\ DFT showing a dominant peak at the spin frequency.}
\label{o:dippsp}
\end{center}
\end{figure}
\cite{buc92} reported on an unusual and sudden temporary drop in flux of EX Hya in decline, which does not coincide with the eclipse time.  This is illustrated in Figure~\ref{o:diplcfit} for the \heo\ \el\ (\hb, \hg\ and the continuum -not shown- show similar results).  \cite{buc92} used the orbital and spin ephemerides of \cite{bon88} to phase their observations, and we used those of \cite{hel92} to phase ours, and our results predict similar time for the occurance of the dip.  The dip is a real event since the observations were conducted under clear, photometric conditions, and the dip was proven not to be due to telescope drive or guiding errors.
 
The dip lasts some 30 minutes (from roughly the time of ingress to that of egress).  This dip is reminiscent of that observed in the previous studies of EX Hya, where binary phased light curves show a bulge eclipse which extends over $\phi_{98}\sim0.6-1.0$ \citep{hur97,mau99,bel02} in the EUV, and around $\phi_{98}\sim0.7-0.1$ \citep{cor85,ros88} in the X-rays.  The bulge is interpreted as an absorption of the EUV emitting region on the \wtd\ surface in the EUV photometry.  In the optical, extended emission at the outer edge of the ring in the \ha\ tomogram was observed by \cite{bel05}, while absorption of the zero-velocity component at $\phi_{98}\sim0.6-0.9$ was observed in \ha, \hb\ and \hg\ \citep{bel05};(see also \cite{mhl06}). 
This was interpreted by \cite{bel02,bel05} as an indication of the existence of an extended bulge near the outer edge of the ring, from where the extended accretion curtains pull material directly, thereby producing a broad bulge dip in the EUV and the absorption in the optical.  
The only difference is that the phasing of the structure reported here is reversed ($\phi_{98}\sim0.1-0.6$).

From our observations we make the following deductions and interpretation for the dip.

(i) The fact that the `arc emission' seen in the \heo\ orbital tomogram in outburst is always redshifted suggests that the material responsible for this emission is moving away from the observer throughout the orbital phase.  This also means the responsible structure/material is changing locations (i.e. always on the side of the observer all around the orbit) rather than being static as in the case of the usual bulge that is normally observed near the bright spot region as discussed above.

(ii) The `arc emission' in outburst extends over spin and binary phases $\phi_{67,98}\sim0.1-0.6$.  This phasing and the physical extent of the emission are in agreement with those of the dip eclipse observed in decline (Figure~\ref{o:diplcfit}).  Therefore to explain the dip, there has to be extended emission in the \heo\ tomogram in decline similar to that seen in outburst, and the reason we do not observed this emission could be that the responsible material/structure has a high opacity in decline, just as in the EUV case.   

\begin{figure}
\begin{center}
\includegraphics[width=80mm]{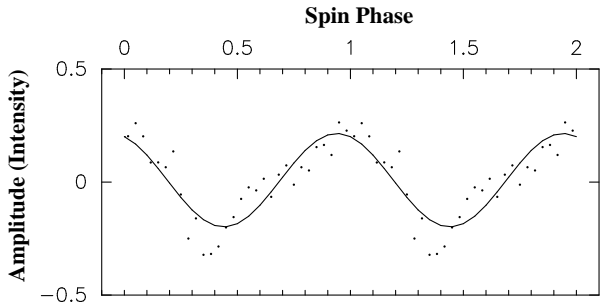}
\caption{\heo\ emission phase folded on the spin cycle.  The dip shows minimum flux near phase 0.4. The solid line represents a fit to the data.}
\label{o:dipfld}
\end{center}
\end{figure}

(iii) In decline the data are strongly modulated at the spin period (Figures~\ref{o:diplcfit}, \ref{o:dippsp} and \ref{o:dipfld} - see also \rv\ DFTs in Figure~\ref{rv:2891hbg-s}) and the fact that the phase of the dip minimum (0.4) nearly corresponds with spin minimum (when the upper accretion curtain is pointing at the observer) suggests that the accretion curtain and its location is strongly related with the structure that is causing the dip.

(iv) The dip minimum is seen at phase 0.4 in the orbital phase as well.  At this phase the observer is behind the material causing the `arc emission'.  This is a phase where spin minimum is observed (at 0.4), i.e. when the upper accretion curtain is on the side of the observer (between the observer and the \wtd).  An optically thick accretion curtain can obscure \wtd\ emission at this phase.

(v) The fact that the flux at spin pulse maximum (just before the dip) is less than that in the earlier high points in the light curve, and that the dip is deeper than at the earlier low points (spin minima), suggests that there is extra absorption to that caused by photoelectric effect up/down and across the accretion curtains (maybe combined with that caused by occultation of one of the poles).

We therefore suggest that the material responsible for the `arc emission' in outburst is optically thick in decline and obscures the accretion curtain and the \wtd\ emission (from both the continuum and the emission lines).  This material is probably picked/swept up by the field lines from the bulge or from the ring (as the magnetospheric radius increases in decline) and corotates with the extended accretion curtain, as was first observed in \cite{mhl06}.  The absorption caused by this material combines with the normal photoelectric absorption, resulting in almost all the light from the compact object and the accretion curtains being obscured, from phases $\phi_{98}=0.1-0.6$.  The fact that the dip lasts some 30 mins suggests that this material and the accretion curtains extend to nearly half the accretion ring.  

Figure~\ref{o:dipfld} shows the data after phase-folding on the spin ephemeris of \cite{hel92}. Maximum is seen near $\phi\sim0.9-1.0$, in agreement with the accretion curtain model.

The drop in maximum intensity nearly half a spin cycle before the dip (Figures~\ref{o:diplcfit}) could therefore suggest that the optical depth of the accretion curtains increased at that moment as the optically thick material was being picked up by the magnetic field lines of the \wtd. 

The structure causing the dip also absorbs zero velocity emission.
\cite{buc92} reported that the \heo\ central absorption gets stronger before the dip occurs.  This further establishes a connection between the dip and the absorption of the zero velocity emission in \heo\ during decline.  

The fact that this event is not observed to repeat could suggest variations in the optical depth of the bulge associated only with outburst or with the decline phase.  
\section{Discussion}
\label{sec:odisc}
In this section we compare disc instabilities with mass transfer events and decide, on the basis of our observations, whether outbursts in EX Hya occur via one of the two mechanisms or both or any variation on the two.

According to the generally accepted model, dwarf novae outbursts are caused by an instability in the disc, resulting in the rapid accretion of material onto the surface of the \wtd.  This notion is supported by the argument that the observed flux from the bright spot is similar in quiescence and in outburst.  The disc instability model has had success in describing outbursts in dwarf novae.  It is thought that the same mechanism may drive outbursts in IPs even though their inner discs are truncated as a result of the relatively strong magnetic field.  If outbursts occur via a disc instability, then the fact that EX Hya accretes from a ring of material and the accretion curtains extend to this ring could suggest a different instability, perhaps similar to that described by \cite{spr93}, where outburst cycles result due to storage and release of matter in the parts of the disc just outside the magnetosphere.  

Alternative ideas have outbursts caused by a \snd\ star instability or a burst of mass transfer from the \snd.  In at least two IPs, TV Col and EX Hya, the observations of a dramatic increase in the S-wave flux and of overflow stream (EX Hya) in outburst have been taken as evidence for the occurence of a mass transfer burst \citep{hel89b, hel93}.  \cite{hel89b} claimed that the equivalent width of the S-wave of the \el\ was similar in outburst and in quiescence during the time of their observations, and concluded that the flux from the S-wave increased during the 1987 outburst.  They saw this as evidence for an increased mass transfer, but it was not conclusive.
There are `hybrid models' which seek to combine disc instabilities with mass transfer events to explain outbursts in SU UMa stars and in most low mass X-ray binaries (LMXB), where it is thought that enhanced mass transfer may generate a disc instability, causing an outburst \citep{lny96}.  A similar but reversed mechanism has already been suggested as an alternative for EX Hya \citep{hel00a}.

In investigating the feasibility of the mass transfer burst models in EX Hya, first we note that our spectroscopy provides evidence for strong irradiation of the \snd\ star and the inner disc in outburst.  
All the \els\ have shown strong emission from the \snd\ Roche lobe.  The \het\ tomograms, on the one hand, have revealed a larger ring of emission (in velocity) when compared to those of other emission lines in outburst (Figure~\ref{o:he12idltrl}).  On the contrary, \hb, \hg\ and \heo, showed significantly and relatively less inner disc emission and much more from the core during outburst than in quiescence.  Correspondingly, the line profile widths have shown a significant decrease in \hb, \hg\ and \heo\ during outburst, by nearly 33\%-55\% when compared to quiescence (Figure~\ref{s:superimp}).  
This suggests irradiation of the inner disc which is possible if the ionization flux due to the UV and soft X-ray emission from the \wtd\ was extremely high, as expected in outburst.  In this situation, recombination will be relatively reduced most of the time, resulting in the decrease of line emission or absorption, except for \het, which is irradiation-driven.
Furthermore, the \het\ tomogram shows an asymmetry on the `far' side, suggesting that most of the disc light comes from this side, which is not eclipsed by the \snd\ (Coel Hellier; private communication). 

Observations in decline show an increase in the widths of the emission lines by $\sim$20\% (\hb\ and \hg) (Figure~\ref{s:superimp}) and by $\sim$33\% (\heo).  Correspondingly, strong disc emission seen at low velocities in outburst moves to higher velocities during decline, indicating that relatively more emission was received during this period.  
In addition, relatively more emission from near the bright spot, and relatively less from the \snd, were observed.  All this is an indication that the irradiation of the \snd\ and that of the disc occured in outburst and had stopped or decreased during decline.

Secondly, we think that the irradiation of the front face of the \snd\ (and evidence for disc overflow - discussed later in the Section) is not a strong indication of a \snd\ instability or an increase in mass transfer.
The disc instability model predicts a constant S-wave which is swamped in outburst, while the mass transfer burst model predicts that the bright spot flux will contribute to outburst.  
The latter model further requires a mass transfer burst which is observationally estimated as $\sim2\times10^{22}$ g in EX Hya \citep{hel93}.  This implies a mass transfer rate of $\sim8\times10^{17}$ g s$^{-1}$ (averaged over a period of $\sim6$ hrs near the peak of outburst) and a bright spot luminosity of roughly 

\begin{equation}
L_{sp,outb}\sim\frac{GM_{WD} \dot{M}(2)}{R_{sp}}\sim2\times10^{33} {\rm erg \hspace{0.1cm} s^{-1}},
\label{eq:lspot}
\end{equation} 
which is higher than the quiescent accretion-induced luminosity of the \wtd\ of
\begin{equation}
L_{acc,q}\sim4\pi d^{2} f_{q}=3\times10^{32} {\rm erg \hspace{0.1cm} s^{-1}}
\end{equation}
(at a distance of 65 pc \citep{beu03}) by nearly a factor of 10.  $M_{WD}$ and \.M(2) are the mass of the \wtd\ and the mass transfer rate from the \snd, respectively, and $R_{sp}$ is the radius of the bright spot.   $L_{sp,outb}$ is higher than the outburst accretion-induced luminosity
\begin{equation}
L_{acc,outb}=\frac{GM_{WD} \dot{M}}{2R_{WD}}=2\times10^{34} {\rm erg \hspace{0.1cm} s^{-1}},
\label{eq:oacc}
\end{equation} 
where $\dot{M}\sim5\times10^{17}$ g s$^{-1}$ is the mass accretion rate (averaged near the peak of outburst, with the mass involved in outburst taken to be $\sim$10$^{22}$ g - \cite{hel00a}).
For the mass transfer burst idea we would expect $L_{acc,q}<L_{sp,outb}$, as the calculations above confirm (if the numbers were very reliable) since the emission at the bright spot due to the mass transfer burst should be stronger than that from other components in the system.   But this is not supported by our observations because the bright spot, which is clearly present in quiescence and in decline, is not observed in outburst, suggesting that it is swamped in outburst.  Also, as shown above, $L_{sp,outb}<L_{acc,outb}$.  This is not expected in a mass transfer burst situation and this is what our observations seem to support.  This observation and arguments counts against the mass transfer burst model.

Also, iradiation of the \snd\ will not be effective enough to trigger a greatly increased mass transfer since the height, h$_{d}$, of the outer edge of the ring, 
 
\begin{equation}
h_{d}\sim0.038r_{d}\dot{M(d)}^{\frac{3}{20}}_{16}\sim10^{9} {\rm cm},
\label{eq:heightd}
\end{equation}
for a 0.5 M$_{\odot}$ \wtd\ and for a mass transfer rate $\dot{M}(d)$ in the disc of 0.8$\times10^{16}$ g s$^{-1}$ \citep{war95}
is higher than the scale height,
\begin{equation}
H_{L1}\sim1.5\times10^{8} {\rm cm},
\end{equation}
of a gas column at L$_{1}$ \citep{mey83,kol90}.
This suggests that the shadow of the disc will reduce the irradiation of the surface of the \snd\ near L$_{1}$.
(\cite{eis02} predicted a disc thickness of 2$\times10^8$ cm and a disc radius of 1.6$\times10^{10}$ cm.  The latter value is less by nearly a factor of two when compared to that of $\sim$3$\times10^{10}$ cm obtained by Mhlahlo et al. (2006) and \cite{hel87}.  When \cite{eis02}'s value of disc radius is put in Equation \ref{eq:heightd} then a higher value for $h_{d}$ ($\sim6\times10^{8}$ cm) than they predicted is obtained, and so our result obtained from Equation \ref{eq:heightd} above is being considered here).

It is important to point out, though, that the formulae used here are for simple steady state truncated discs and may not necessarily apply properly to EX Hya.
But according to our quiescence model of EX Hya \citep{mhl06,bel05,bel02,kin99,nor04}, the ring of material at the outer edge is disrupted by the field.  This should result in the effective disc height at the rim being higher than the value given above since the magnetic field would increase the density of material at the outer disc edge ($h_{d}$ depends on the temperature, density and local gravity of material).  
Observational evidence for a bulge in EX Hya at the outer edge of the disc \citep{mau99,bel02,bel05,hoo05} and corotating raised material (Section~\ref{sec:dips}) supports this claim.  This bulge is raised by the accretion curtain for nearly half the orbital cycle and so will shadow the L$_{1}$ point below and above the disc for nearly this period during every orbital cycle.  
It is therefore unlikely that outbursts are due to a mass transfer event from the irradiation of the \snd\ surface.

Furthermore, the accretion-induced radiation in outburst (Equation~\ref{eq:oacc}), will not reach deep into the \snd's photosphere, since it lies in the extreme ultraviolet and soft X-ray regions where opacities are very large.  This radiation will be absorbed by photoionisation \citep{hkl86} and reradiated in high layers of the \snd's hemisphere, producing strong chromospheric emission.  The same emission will irradiate the inner disc and the accretion curtains, giving rise to strong \het\ emission and to suppression of the Balmer lines which we observe in our data.  The radiation in quiescence and that resulting from outburst therefore will not greatly affect the mass transfer rate, other than increasing chromospheric emission.  The same is true for the hard X-ray emisssion in outburst.

Most of our data fit the mechanism of \cite{spr93}.  \cite{spr93} showed that conditions at the inner edge of the disc can cause variations of the magnetosphere boundary, and this in turn will result in the accretion near the value of the fastness parameter, $\omega_{s}=1$ (see Section~\ref{sec:omod} for a detailed discussion) being unstable.  They showed that matter accumulates outside the magnetosphere and, when the gas density in the disc exceeds a critical value, forces the magnetospheric boundary inwards, resulting in reduced value of $\omega_{s}$ ($\omega_{s}<1$).  The latter results in the density at the magnetospheric boundary dropping, and the consequent density gradient at the magnetospheric radius causing an increase in the mass flow.  Matter is then forced to accrete and its depletion leads to the surface density of the magnetosphere boundary decreasing.  The magnetospheric boundary consequently is forced outward to radii corresponding to $\omega_{s}>1$ at a reduced accretion rate, sweeping up material from the disc and forcing it to corotate with it.  When sufficient mass has piled up at the inner region of the accretion disc it pushes the magnetosphere back to $\omega_{s}<1$, and the cycle repeats nearly on a viscous time scale.  \cite{spr93} pointed out that their model could be applied to IPs.  This was taken up by \cite{war96} who suggested that this model can be applied to EX Hya.

Our data fit this mechanism as follows:

(i) Measurements of peak-to-peak separation of $\sim700$ \ks\ (estimated from a double-gaussian fit of the \hb\ \el) have indicated a rotational velocity of $\sim350$ \ks, leading to a radial distance from the \wtd\ to the outer edge of the ring of $\sim5\times10^{10}$ cm, which is $\sim$ $a$, the binary separation (for a \wtd\ mass of 0.5 M$_{\odot}$; Keplerian motion about the \wtd\ at the outer edge of the disc was assumed).  As discussed in \cite{mhl06}, the disc is disrupted by the field near $\sim3\times10^{10}$ cm and so the ring which extends from this distance to near $a$ has a width of $\sim2\times10^{10}$ cm.

The brightning at the outer edge of this ring is observed in the outburst tomograms, except in \het\ (Figures~\ref{o:hbgidltrl} and \ref{o:he12idltrl}).  In addition, the \heo\ orbital tomograms show an `arc emission' near $\sim350-400$ \ks, covering about half the orbital phase.  This can be accepted as evidence that some material had accumulated at the outer edge of the ring and that there was a greatly extended ring of material in outburst.  

(ii) The orbital trailed spectra have shown redshifted emission which can be associated with the `arc emission' in the tomograms, near the outer ring.  This emission is observed throughout the entire orbital cycle in outburst, while the blueshifted emission is obscured by the accretion curtain, when it is on the side of the observer and when it is furthest from the observer.  This is evidence for the ring material drifting away from the observer (suggesting that the material is drifting towards the field lines of both the lower and the upper poles). This is also evidence for the corotation of some of this material with the accretion curtains (see (i)-(vi) in Section~\ref{sec:dips}).  

(iii) Our discovery of \het, \hb\ and \hg\ emission on the upper accretion curtain at higher velocities than in quiescence, when the \pri\ was between the upper accretion curtain and the \snd\ (Figures~\ref{o:avsubmaps}), provided evidence for disc overflow accretion.  This emission was modulated at $\pm$1500 km s$^{-1}$ whereas in quiescence it was measured at $\pm$1000 km s$^{-1}$, suggesting that the overflow material hit the magnetosphere much closer to the \wtd.
The VHVC was modulated at the orbital period (Figures~\ref{rv:27hbg-as} and~\ref{rv:2791hbghe2-3t7h1t2}) since the accretion curtains pick up material mostly when it is fed favourably from the stream, i.e. on the side facing the magnetic pole.
\begin{figure*}
\begin{center}
\includegraphics[width=110mm]{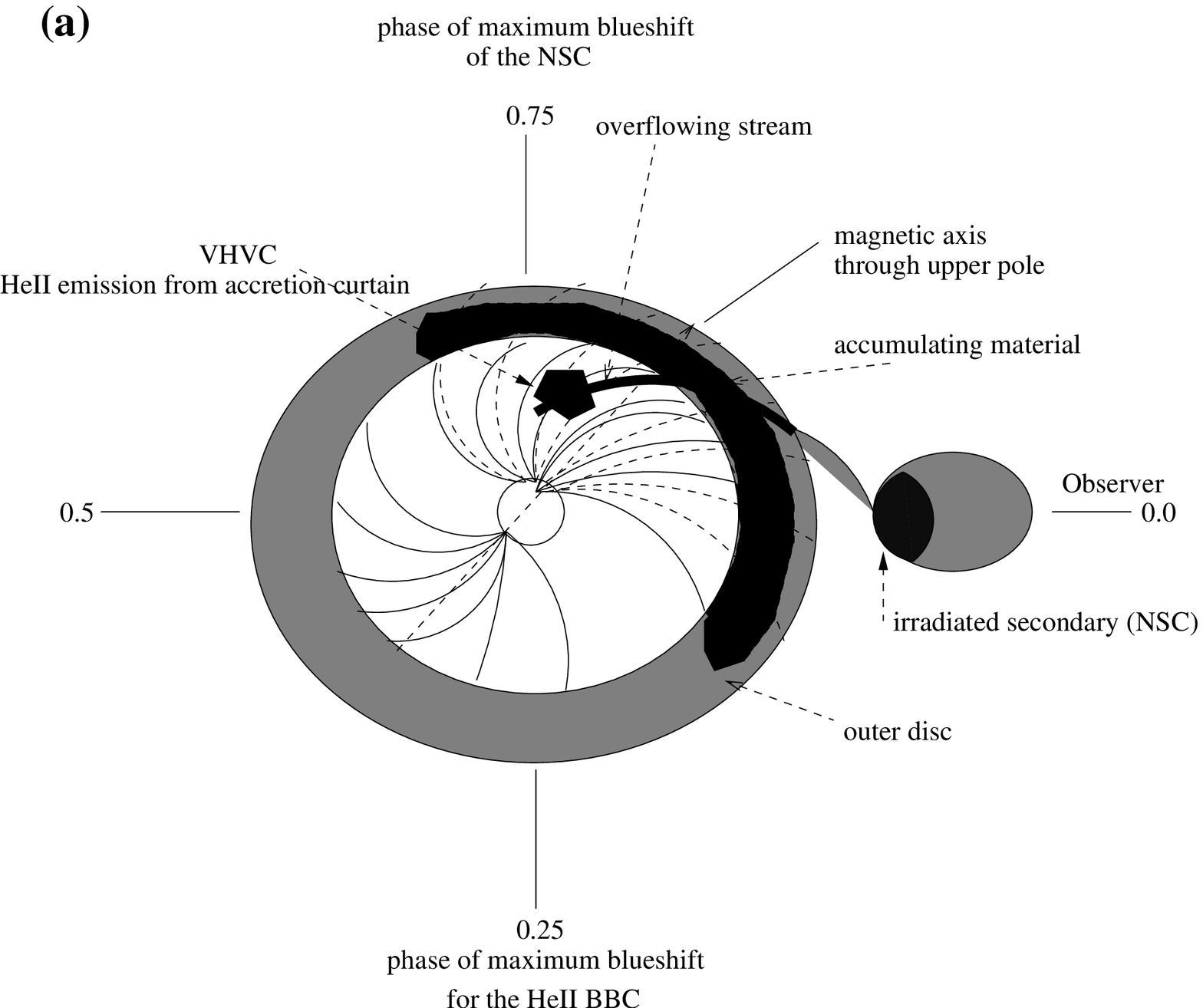}
\includegraphics[width=110mm]{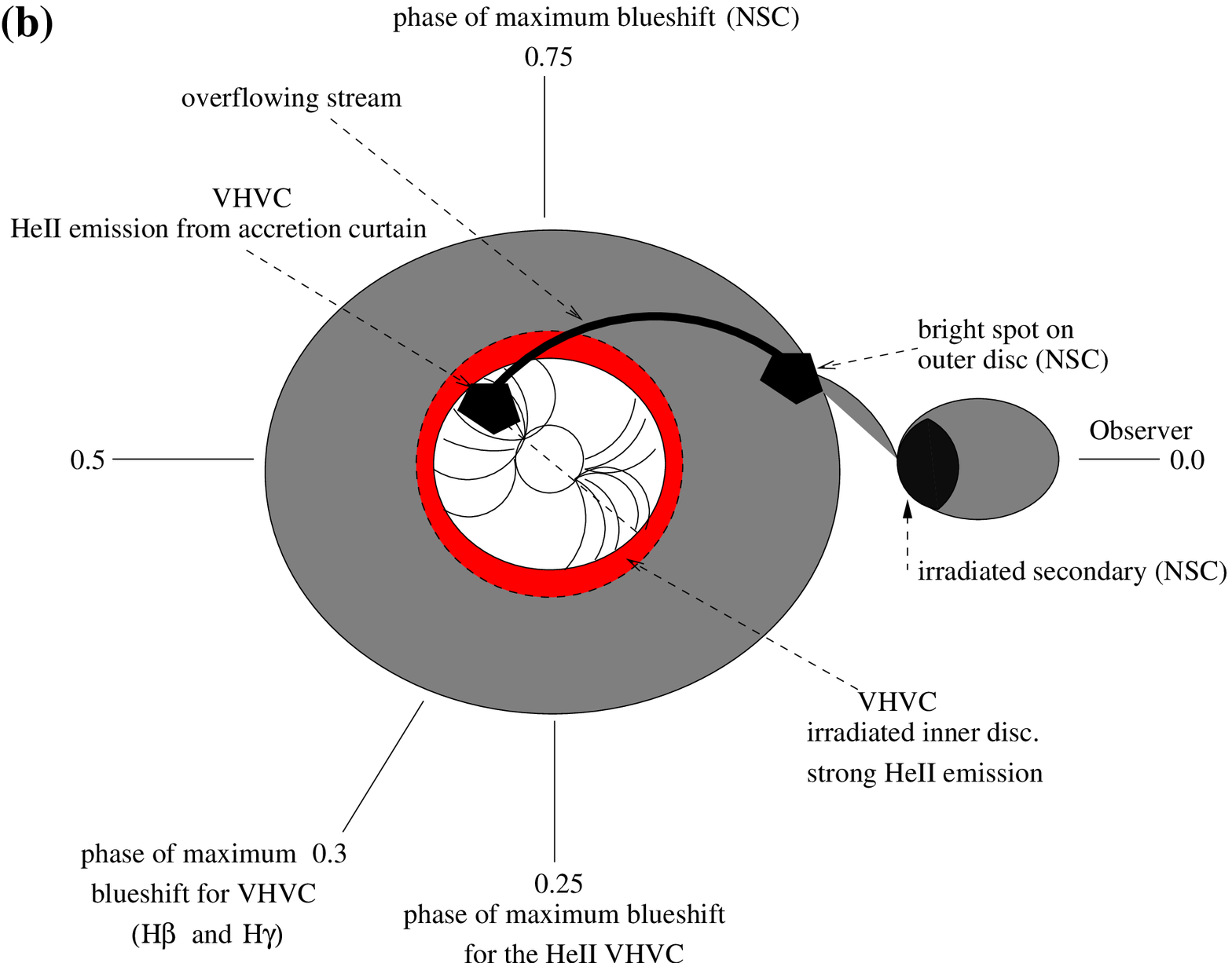}
\caption{\small A schematic drawing of EX Hya just before outburst (a) showing material accumulating near $r_{in}$ and forced to corotate with the magnetosphere at the inner edge of the ring.  Diagram (b) depicts a situation during outburst, after the magnetosphere has been pushed inwards, and in decline before the magnetosphere is forced outwards.  The emission region on the magnetosphere due to impact by the overflowing stream of material is depicted.  The inner regions of the disc emitted relatively less \hb\ and \hg\ but relatively more \het\ emission due to irradiation.  Most of this emission came from the rear side of the disc.  The stream emission hit the magnetosphere closer to the \wtd\ during outburst (VHVC) than in quiescence (HVC).}
\label{o:oexillus1}
\end{center}
\end{figure*} 
The phasing of the VHVC (maximum blueshift at $\phi_{98} \sim 0.3-0.4$) is consistent with the expected phase of impact of a stream with the magnetosphere, and the phase lag between the \hb, \hg\ and \het\ NSC and VHVC of $\sim180^{\circ}$ is consistent with the motion of the region hit by the overflow stream on the magnetosphere (i.e. the region behind the \wtd\ relative to the front face of the \snd) being anti-phased with the motion of the \snd.  
The \rvs\ show large negative values of $\gamma$-velocities in the blue for HeII $\lambda4686$ (-304 \ks\ when compared to -67 \ks\ for \hb) which indicates streaming velocities much closer to the \wtd\ \citep{buc00}.  
In previous studies of EX Hya \citep{hel89b,hel00a}, the observation of the overflow stream during outburst was interpreted as evidence for enhanced mass transfer.  We point out that the fact that the overflow stream was modulated at higher velocities in outburst than in quiescence suggests that the radius of the magnetosphere decreased, and so the overflow material hits the magnetosphere closer to the \wtd\ than when the accretion curtain is extended (Figure~\ref{o:oexillus1}).

(iv) The VHVC which was prominent in outburst was not present in decline, instead, a HVC (near $\sim1000$ \ks) was observed and this, by parity of reasoning (see (iii) above), indicates that the overflow stream hit the magnetosphere far from the \wtd\ as is the case in quiescence \citep{mhl06} due to the magnetospheric boundary having been forced outwards.

\subsubsection{Revised Outburst Model of EX Hya}
\label{sec:omod}
We therefore propose that outbursts in EX Hya are triggered by a mechanism similar to that suggested by \cite{spr93} as illustrated in the rough sketches in Figure~\ref{o:oexillus1}.

If so then this would imply that mass was stored in the outer ring near $\omega_{s}>1$, as suggested by observations, where $\omega_{s}=\frac{\Omega_{s}}{\Omega_{k}(r_{in})}$ is the dimensionless stellar angular velocity or the fastness parameter, ${\Omega_{s}}$ is the spin angular velocity of the \wtd\ and $\Omega_{K}(r_{in})$ is the Keplerian angular velocity at the inner radius $r_{in}$ of the disc.  
Where the inner edge of the ring of EX Hya coincides with the corotation radius,  
$\omega_{s}=0.329\mu(1)_{33}^{6/7}M_{WD}^{-5/7}P_{3,spin}^{-1} \dot{M}^{-3/7}_{17}\sim1$,
assuming a magnetic moment of 5 $\times10^{33}$ \citep{nor04}, and \.{M}$\sim0.1\times10^{17}$ g s$^{-1}$ \citep{fuj97,hel00a}.  $P_{3,spin}$ is in units of 10$^{3}$ s.  At this $\omega_{s}$ the material is moving with a velocity
\begin{equation}
v_{d}(r_{in})=943 \left [\frac{M_{WD}}{\omega_{s}P_{3,spin}} \right]^{\frac{1}{3}}\sim 500 \hspace{0.1cm} {\rm km \hspace{0.1cm} s^{-1}},
\end{equation}
\citep{war95}, where here M$_{WD}$ is in units of M$_{\odot}$ (solar masses), in agreement with our observation in quiescence \citep{bel05,mhl06}. 

The amount of material transfered from the \snd\ to the disc (with a small percentage overflowing the disc) over the recurrence interval ($\sim$1.5 yr) can be estimated to be $\sim1 \times10^{23}$ g (assuming \.M(2)$\sim2\times10^{15}$ g s$^{-1}$ \citep{nor04}), which suggests that over this period of time there will be enough stored material in the ring ($\ge10^{22}$ g - \citep{hel00a}) to trigger an outburst.  

The viscous time scale at r$_{co}$ predicted by this model is $t_{0}=1/\bar{\nu_{0}}\Omega_{s}\sim7$ h, where $\bar{\nu_{0}}=\nu_{0}/(r_{co}^{2}\Omega_{s})\sim0.03$ is a dimensionless viscosity and $\nu_{0}$ is a constant \citep{spr93}.  This is low by nearly a factor of two when compared to the observed rise time in EX Hya of $\le12$ h \citep{hel89b,hel00a}.  

No prediction for the recurrence time is possible using the \cite{spr93} model.  However, it is worth noting that in this model outbursts are due to accumulation of mass, but because there is a lot of mass falling onto the \wtd\ all the time in EX Hya, it takes long to accumulate enough mass to trigger an outburst, leading to a long recurrence period.  Dwarf novae accumulate mass most of the time and so outbursts recur on a time scale of $\sim30$ days to a few months.

\subsection{Summary}

From our analysis we have shown that strong irradiation of the \snd\ star and that of the inner regions of the disc took place in outburst and that most of the \hb\ and \hg\ inner ring and accretion curtain emission was suppressed during outburst.  

 Evidence for the stream of material overflowing the initial impact with the accretion disc and impacting onto the magnetosphere of the \wtd\ star was observed.  In outburst, the overflow material was observed to be modulated at even higher velocities ($\sim1500$ \ks) than in quiescence, which are streaming velocities close to the \wtd.  This suggests a decrease in the magnetospheric radius. 

We showed that the outburst did not occur via the \.M instability nor was there any increased mass transfer due to irradiation of the front face of the \snd.  It is clear that the strong irradiation of the front face of the \snd\ resulted in chromospheric emission.  We suggest that the amount of matter stored in the outer ring is enough to trigger a disc instability of the form suggested by \cite{spr93}.  We explained the dip during decline as due to the extended accretion curtains and the corotating material near the Roche lobe radius obscuring the emission from the \wtd.

Comparison studies of EX Hya and the EX Hya-like system, V1025 Cen, in quiescence and in outburst, will increase our understanding of accretion dynamics and processes in these systems, and further confirm the extended accretion curtain theory.

\section*{Acknowledgments}

{NM would like to acknowledge financial support from the Sainsbury/Linsbury Fellowship Trust and the University of Cape Town.  NM would also like to thank Frank Bateson and W. Allen of the Variable Star Section of the Royal Astronomical Society of New Zealand for supplying visual observations.  We would also like to thank Kunegunda Belle and Coel Hellier for invaluable discussions and for their constructive comments.  We acknowledge use of D. O'Donoghue's and Tom Marsh's programs Eagle and Molly, respectively.  BW is supported by research funds from the University of Cape Town.}

\addcontentsline{toc}{chapter}{References}
\renewcommand{\bibname}{References}

\end{document}